\documentclass[baaa]{baaa}

\usepackage[pdftex]{hyperref}
\usepackage{subfigure}
\usepackage{natbib}
\usepackage{helvet,soul}
\usepackage[font=small]{caption}

\contriblanguage{1}

\contribtype{1}

\thematicarea{6}

\received{\ldots}
\accepted{\ldots}

\title{The distribution of metals in  simulated \\ Milky Way-mass galaxies}

\titlerunning{Distribution of metals in Milky Way-type galaxies}

\author{
F.G. Iza\inst{1,2},
S.E. Nuza\inst{1,3},
C. Scannapieco\inst{2,3},
L. Biaus\inst{2}
\&
E. Lozano\inst{2}
}

\authorrunning{Iza et al.}

\contact{fiza@iafe.uba.ar}

\institute{
Instituto de Astronomía y Física del Espacio, CONICET--UBA, Argentina
\and
Departamento de Física, Facultad de Ciencias Exactas y Naturales, UBA, Argentina
\and
Consejo Nacional de Investigaciones Científicas y Técnicas, Argentina
}

\resumen{
Las galaxias se forman debido a la continua acreción de gas hacia el centro de halos de materia oscura que, cuando alcanza altas densidades, da lugar al nacimiento de discos estelares.
A través del análisis de la distribución de edades, se encuentra que las estrellas en las regiones internas de los discos se formaron en épocas tempranas, contrastando con las poblaciones más jóvenes que ocupan regiones más externas.
Este esquema, arquetípico en el proceso de formación galáctica, es conocido como escenario ``inside-out'' y está íntimamente relacionado con las propiedades globales de una fracción significativa de la población de galaxias espirales.
En este estudio analizamos galaxias del proyecto Auriga, un conjunto de simulaciones galácticas de alta resolución con masas similares a la de la Vía Láctea, para estudiar la distribución de metales a $z=0$.
Dado que cada componente galáctica está sujeta a distintas historias de formación, nos enfocamos en la distribución de metales en el halo, bulbo y disco utilizando un método de descomposición dinámica basado en el grado de rotación de las estrellas y su potencial gravitatorio.
}

\abstract{
Galaxies form due to the continuous accretion of gaseous material towards the centre of dark matter haloes, which gives rise to stellar discs once the gas reaches high density.
Analysis of stellar age distribution shows that stars in the inner regions of the disc are born at earlier times, in contrast to those found in the outer regions, which constitute a younger population.
This scheme, typical of galactic formation processes, is known as the ``inside-out'' scenario and is closely related to the global properties of a significant fraction of the population of spiral galaxies.
In this study, we analyse model galaxies from the Auriga project, a set of high-resolution, magnetohydrodynamic cosmological simulations of Milky Way-mass galaxies, to investigate the distribution of metals at $z=0$.
Since different galactic components were subjected to different formation histories, we focus on the distribution of metals in the halo, bulge, and disc using a dynamical decomposition method based on the degree of rotation of the stars and their gravitational potential.
}

\keywords{galaxies: evolution --- galaxies: structure --- methods: numerical}

\begin{document}

\maketitle
\section{Introduction} \label{sec:introduction}

\begin{figure*}[!t]
    \centering
    \includegraphics[scale=0.9]{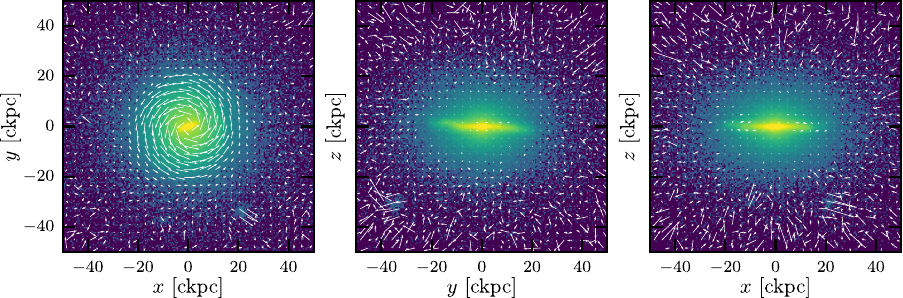}
    \caption{
    Projected stellar density maps for Au9 at $z = 0$.
    The panel on the left shows the face-on view and the middle and right panels the edge-on views.
    The colour map spans five orders of magnitude in projected density using a logarithmic scale.
    The white arrows show the associated velocity field.
    The first `c' in the units stands for comoving coordinates.
    }
    \label{fig:density_map}
\end{figure*}

Galaxies form at the centre of dark matter haloes, which grow from small density fluctuations in the early universe.
In this picture, dark haloes grow hierarchically, both via smooth accretion of matter from the intergalactic medium and  through collisions and mergers with smaller structures.
Initially, the baryonic matter in galaxies consists primarily of hydrogen and helium, the primordial elements of the universe. As stars evolve and reach the end of their life cycles, new, heavier elements are synthesised in their interiors and injected into the interstellar medium (ISM) on varying timescales. However, mergers can complicate the picture by introducing additional chemical elements with different formation histories during the lifetime of the systems \citep[e.g.,][]{Monachesi2019}.

As a result, the chemical abundances and distribution of different elements can provide valuable insight into the formation and evolution history of galaxies. Furthermore, since various galactic components have been found to form at characteristic times, they are also expected to host different stellar populations \citep[e.g.,][]{Matteucci2019, Matteucci2021}.
Spheroids, which are typically formed at earlier times, are expected to be composed of old stars with high levels of $\alpha$-elements and low levels of iron. Conversely, galactic discs are thought to be formed after the first several billion years of evolution and are expected to contain younger stars displaying low $\alpha$-element abundances but enriched in iron-peak elements.

In this work, we study the present-day ($z=0$) distribution of metals in the different galactic components of simulated Milky Way-mass galaxies from the Auriga Project \citep{Grand2017}, a set of \emph{zoom-in} simulations computed with the magnetohydrodynamic (MHD) and cosmological code {\scshape arepo} \citep{Springel2010}.
For each galaxy, we tag stars according to galactic component using a simple dynamical decomposition based on the rotation state and the gravitational potential of each star. For the different components, we then study the distribution of stellar ages, [Fe/H] and [O/Fe] metal abundances.

The organisation of this work is as follows.
In Sect.~\ref{sec:auriga} we describe the main properties of galaxies belonging to the Auriga Project and the selected sample of galaxies.
In Sect.~\ref{sec:decomposition} we present the details of the dynamical decomposition used to tag stars in three distinct galactic components: halo, bulge, and disc.
In Sect.~\ref{sec:amr} we show the age-metallicity relation for Au9, a reference galaxy, and for the sample of selected galaxies.
Sect.~\ref{sec:abundances} describes the relation found between the [O/Fe] and [Fe/H] abundances of the sample.
Finally, in Sect.~\ref{sec:summary} we present a summary of our findings.

\section{The Auriga Project} \label{sec:auriga}

The Auriga Project \citep{Grand2017} is a suite of 30 high-resolution Milky Way-like galaxies simulated in a cosmological environment using the magnetohydrodynamic (MHD) code \textsc{arepo} \citep{Springel2010}, a quasi-Lagrangian, dynamic mesh code that tracks the evolution of MHD and collisionless dynamics in a full cosmological framework.
All simulations follow the evolution of matter from $z = 127$ to $z = 0$ in a $\Lambda$CDM universe.
Among the many galaxy formation processes included in the simulations are primordial and metal-line cooling, a uniform ultraviolet background field, star formation based on gas density, magnetic fields, active galactic nuclei, and chemical feedback from Type Ia and II supernovae and asymptotic giant branch stars.

The simulated galaxies have $z = 0$ virial masses in the range $\sim 9$--$17 \times 10^{11} ~ \mathrm{M}_\odot$ and stellar masses in the range $\sim 3$--$12 \times 10^{10} ~ \mathrm{M}_\odot$.
Furthermore, most galaxies in the sample have developed stable discs during their formation, with disc-to-total mass fractions as high as 0.88 and disc radii that range between 7.9 and $33.7~\mathrm{kpc}$ at $z=0$ \citep{Iza2022}.

Of the 30 galaxies of the Auriga project, we consider only a subset of galaxies that have been subject to a smooth evolution and have well-developed discs in the present \citep{Iza2022, Iza2024}.
Each of these galaxies is labelled as ``Au'', followed by an integer.

In Fig.~\ref{fig:density_map} we show the face-on and edge-on distributions of stars for galaxy Au9.
The figure shows a density map that spans five orders of magnitude in projected density using a logarithmic scale and the associated velocity field.
Au9 is a galaxy with a quiescent formation history, which has not had major mergers in the last several billion years, and which has a well-developed disc at $z=0$.
In this sense, Au9 reproduces the typical behaviour of the sample and, hence, we consider it a reference galaxy.

\section{The Galactic Decomposition} \label{sec:decomposition}

In order to describe the behaviour of the different galactic components, we classify stars based on a simple dynamical method using two parameters: the circularity parameter $\epsilon$ \citep[e.g.,][]{Scannapieco2009} and the normalised gravitational potential $\tilde{e}$. This selection procedure is similar to the method followed by \cite{Du2020}.

After aligning the galaxy in such a way that the stellar disc is contained in the $xy$ plane, we define the circularity parameter for each particle as the ratio of the specific angular momentum in the $z$-direction to the angular momentum of a circular orbit of the radius of the particle (the circular angular momentum): $\epsilon = j_z j_\mathrm{circ}^{-1}$, where $j_\mathrm{circ} = r v_\mathrm{circ}(r)$, with $v_\mathrm{circ}(r) = \sqrt{G M(r) / r}$ and $M(r)$ the mass contained inside a sphere of radius $r$.
With this definition, stars with near-circular orbits have $\epsilon \approx 1$ and those with random, dispersion-dominated orbits have $\epsilon \approx 0$.

The normalised potential for each star is the gravitational potential $e$ normalised to the maximum value: $\tilde{e} = e \left| e \right|^{-1}_{\rm max}$.
This number is constrained between -1 and 0 and indicates how gravitationally bound a given star is.
A particle with $\tilde{e} \approx -1$ is located near the centre of the potential well, while $\tilde{e} \approx 0$ represents a star far away from the galactic centre.

Using the parameters $\epsilon_\mathrm{rot} = 0.4$ (the threshold to separate rotating from non-rotating components) and $\tilde{e}_0 = -0.6$ (the threshold between the most and least bound stellar structures), we define the galactic components as follows:

\begin{itemize}
    \item The \emph{halo} consists of non-rotating ($\epsilon < \epsilon_\mathrm{rot}$) and less bound ($\tilde{e} > \tilde{e}_0$) stars.
    \item The \emph{bulge} consists of non-rotating ($\epsilon < \epsilon_\mathrm{rot}$) and more bound ($\tilde{e} \leq \tilde{e}_0$) stars.
    \item The \emph{disc} consists of rotating stars ($\epsilon \geq \epsilon_\mathrm{rot}$).
\end{itemize}

\begin{figure}[!t]
    \centering
    \includegraphics[width=0.8\columnwidth]{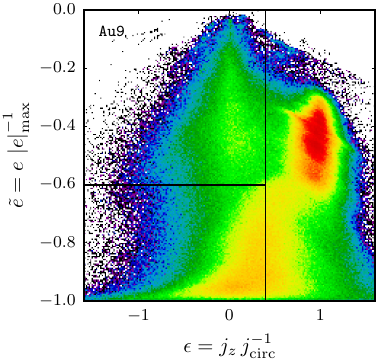}
    \caption{
    Distribution of stars in the normalised gravitational potential-circularity parameter space for Au9 at $z = 0$.
    The solid lines indicate the separation between different galactic components, as described in the text.
    }
    \label{fig:decomposition}
\end{figure}

Fig.~\ref{fig:decomposition} shows the distribution of stars in the decomposition plane ($\epsilon$,$\tilde{e}$) for Au9, along with lines that indicate the limit of the galactic components as previously defined.
Since the figure uses a logarithmic colour map for particle counts, it is clear that the vast majority of the stars of the galaxy inhabit the disc following an ordered rotation with $\epsilon \approx 1$.

\section{The Age-Metallicity Relation} \label{sec:amr}

\begin{figure*}[!t]
    \centering
    \includegraphics{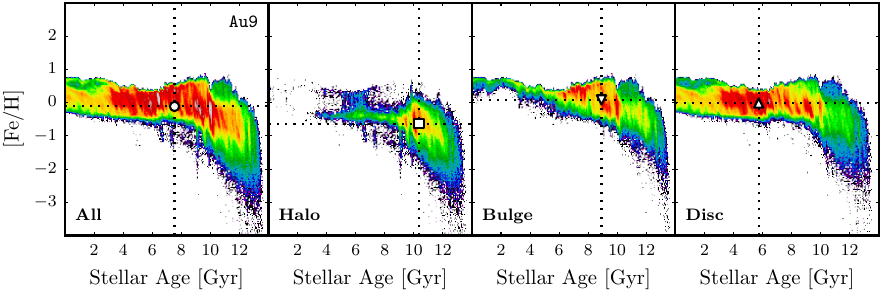}
    \caption{
    Age-metallicity relation for the stars of Au9 at $z=0$.
    In the $x$-axis we indicate the stellar age in Gyr and in the $y$-axis we show the [Fe/H] abundance ratio.
    The colour map indicates the amount of stars in each pixel in a logarithmic scale.
    Each panel shows the distribution of stars belonging to a specific galactic component.
    \emph{First panel:} Age-metallicity relation for all stars.
    \emph{Second panel:} Age-metallicity relation for halo stars.
    \emph{Third panel:} Age-metallicity relation for bulge stars.
    \emph{Fourth panel:} Age-metallicity relation for disc stars.
    In each panel, the symbol indicates the location of the median.
    }
    \label{fig:amr}
\end{figure*}

\begin{figure*}[!t]
    \centering
    \includegraphics{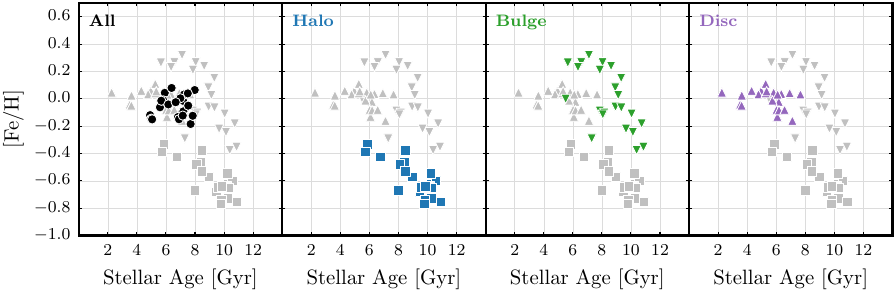}
    \caption{
    Median [Fe/H] abundance versus median stellar age at $z=0$.
    Each point in these scatter plots is a galaxy and the grey symbols in the background show the values for all components.
    \emph{First panel:} [Fe/H] versus stellar age for all stars.
    \emph{Second panel:} [Fe/H] versus stellar age for halo stars.
    \emph{Third panel:} [Fe/H] versus stellar age for bulge stars.
    \emph{Fourth panel:} [Fe/H] versus stellar age for disc stars.
    }
    \label{fig:iron_vs_age}
\end{figure*}

In Fig.~\ref{fig:amr} we show the so-called age-metallicity relation (AMR) for the stars of Au9.
Each panel shows the distribution of stars in the [Fe/H] metal abundance ratio versus the stellar age; from left to right, we show the distribution of a specific galactic component: all stars, halo stars, bulge stars, and disc stars.
Fig.~\ref{fig:amr} shows a well-defined, negative correlation between the [Fe/H] abundance and the age of the stars in the galaxy with a sharper slope and a wider deviation for older stars.
This sharp increase of the [Fe/H] abundance at early times (older stars) is due to a burst of star formation that takes place during the first $4~\mathrm{Gyr}$ of evolution and that reaches a maximum SFR of about $10~\mathrm{M}_\odot \, \mathrm{yr}^{-1}$. This burst, concurrent with the collapse of the gas, provides a rapid enrichment of the interstellar medium (ISM) due to the feedback processes associated with supernovae.

Spheroidal components tend to be populated by older stars, with median stellar ages around $\sim 9$--$10~\mathrm{Gyr}$ for the bulge and halo, while the disc has median ages of $\sim 6~\mathrm{Gyr}$.
In terms of the [Fe/H] abundance, discs tend to be more metallic, with near-Solar values for the median, while the haloes are less metallic, with sub-Solar metallicities.
Bulges have [Fe/H] abundances similar to that of the discs but with greater spread.

Furthermore, in Fig.~\ref{fig:iron_vs_age} we show the age-metallicity relation for all galaxies in the sample, where each data point corresponds to their median values.
In general terms, there is a clear distinction of the stellar population in each galactic component.
In our sample of galaxies, haloes are located in the low metallicity (with a distribution peaking at $[\mathrm{Fe}/\mathrm{H}]\sim -0.7$), high stellar age ($\sim 10~\mathrm{Gyr}$) corner of the diagram, while discs are populated by more metallic (peaking at Solar values) and younger stars (with typical stellar ages of $\sim 5$--$7~\mathrm{Gyr}$).
The bulge presents a somewhat intermediate population between that of the halo and the disc.

\section{Metal Abundance Ratios} \label{sec:abundances}

\begin{figure*}[!t]
    \centering
    \includegraphics{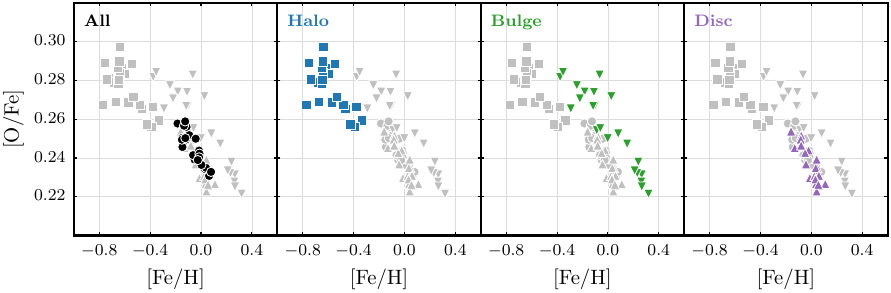}
    \caption{
    Median [O/Fe] abundance versus median [Fe/H] abundance at $z=0$.
    Each point in these scatter plots is a galaxy and the grey dots in the background show the values for all components.
    \emph{First panel:} [O/Fe] versus [Fe/H] for all stars.
    \emph{Second panel:} [O/Fe] versus [Fe/H] for halo stars.
    \emph{Third panel:} [O/Fe] versus [Fe/H] for bulge stars.
    \emph{Fourth panel:} [O/Fe] versus [Fe/H] for disc stars.
    }
    \label{fig:oxygen_vs_iron}
\end{figure*}

Fig.~\ref{fig:oxygen_vs_iron} shows the correlation between the [O/Fe] and the [Fe/H] abundances at $z=0$.
Oxygen, an $\alpha$-process element, is produced mainly by SN\,II while iron is produced mainly by SN\,Ia.
Given that $\alpha$-elements are produced in shorter timescales than iron-peak elements, this figure can be interpreted as a cosmic clock from the upper left corner (higher [O/Fe], lower [Fe/H]) to the bottom right corner (lower [O/Fe], higher [Fe/H]).

By inspecting the galactic components, we find that the haloes of the Auriga sample are typically found in the high [O/Fe]-low [Fe/H] region of the figure while discs are located in the opposite region.
This is in concordance with Fig.~\ref{fig:iron_vs_age}, where haloes are typically populated by older stars than the disc.
As before,  the bulge presents somewhat intermediate results between the halo and the disc, which is in agreement with the fact that these components contain stars of intermediate ages.

\section{Summary} \label{sec:summary}

In this work, we presented an analysis of the distribution of metals at $z=0$ in simulations of Milky Way-mass galaxies from the Auriga project.

We found that the components of the Auriga galaxies share a common behaviour: haloes are populated by old, iron-poor stars, discs are populated by young, iron-rich stars, and bulges are composed mainly by intermediate age populations.
When $\alpha$-elements are considered, like oxygen in this case, the halo presents an oxygen-rich population while the disc is populated by stars with less content of oxygen.

In order to characterise the evolution of metals in the Auriga galaxies, the results presented here will be further complemented with an analysis of the origin and evolution of the distribution of stellar ages and metals in a future work.

\begin{acknowledgement}
S.E.N. and C.S. are members of the Carrera del Investigador Científico of CONICET. They acknowledge support from Agencia Nacional de Promoción Científica y Tecnológica through PICT 2021-GRF-TI-00290 and Universidad de Buenos Aires through UBACyT 20020170100129BA. 
\end{acknowledgement}


\bibliographystyle{baaa}
\small
\bibliography{bibliografia}
 
\end{document}